\documentclass[journal]{IEEEtran}
\usepackage[noadjust]{cite}
\usepackage{amsmath,graphicx,cite,booktabs,amssymb,gensymb, color, xcolor}
\usepackage[linesnumbered,lined,ruled]{algorithm2e}
\usepackage{algorithmic}
\usepackage{subfigure}
\usepackage{multirow}
\usepackage[affil-it]{authblk}

\makeatletter
\renewcommand{\@algocf@capt@plain}{above}

\makeatother
\def\bsx{{\boldsymbol{x}}}
\def\bpsi{{\boldsymbol{\psi}}}

\title{Adaptive transform via quantum signal processing: application to signal and image denoising}
\author{Raphael Smith$^{1}$, Adrian Basarab$^{1}$, Bertrand Georgeot$^{2}$, and Denis Kouam\'e$^{1}$}

\affil{
   {\em $^{1}$IRIT UMR CNRS 5505, University of Toulouse, Universit\'e Paul Sabatier, CNRS, France} \\
 \vspace{.1cm}
  {\em  $^{2}$Laboratoire de Physique Th\'eorique, IRSAMC, Universit\'e de Toulouse, CNRS, UPS, France}
 }

\begin{document}

\maketitle

\begin{abstract}
The main scope of this paper is to show how tools from quantum mechanics, in particular the
Schroedinger equation, can be used to construct an adaptive transform suitable for signal and image processing applications. The proposed dictionary is obtained by considering the signal or image as a discrete potential in Schroedinger equation, further used to construct the Hamiltonien operator. In order to illustrate its practical interest in signal and image processing, we provide denoising results in the case of signal-dependent noise, which is the noise type the most adapted to the proposed approach. 
\end{abstract}

\section{Introduction}
\label{sec:intro}

Transforming digital signals or images in other representations than time or space is a common approach in number of applications such as image compression, restoration, reconstruction, denoising or more recently compressed sensing. A large variety of pre-designed dictionaries exists, including frequency (e.g., Fourier or DCT) or time-frequency (e.g., wavelet or curvelet) transforms. Despite very interesting results, such fixed transforms suffer from a lack of generality and are often only adapted to a particular class of images or signals. To overcome this issue, dictionaries learned from a set of training signals or images have been largely explored over the last decade \cite{Aharon06}. They have been shown to be very efficient in sparsity driven applications, given their potential in sparsifying signals or images (e.g., image denoising \cite{Elad06}).

In this paper, we propose a new way of generating an adaptive basis from the signal or image itself (without the need of a training set as it is usually the case in dictionary learning), by exploiting principles of quantum mechanics. Note that several attempts of translating quantum principles in image or signal processing applications already exist in the literature, such as the seminal work in \cite{eldar2002quantum}. More recently, the interest of quantum mechanics in object extraction from optical images has been shown in \cite{Gabbouj2013} and in pulse-shaped signal analysis in \cite{Laleg2013}. An alternative literature on designing image processing algorithms adapted to quantum computers also exists, but is out of the scope of this paper.  

The starting point of our framework is a discrete version of the Schroedinger equation for a quantum particle in a potential, where the potential is represented by the signal samples or the pixel values. We use a basis of wave functions, \textit{i.e.}, stationary solutions of the Schroedinger equation, as adaptive basis for signal or image representation. Its interest is that the quantum basis automatically uses higher frequencies to describe low potential values. We therefore show that it is well suited to signal-dependent noise (e.g., Poisson), where low signal values need to keep high harmonics, while higher values need more denoising. We note that there was a recent attempt in using quantum mechanics in this context in \cite{LalegKirati15,LalegKirati16}. Although there are similarities between the two approaches, our method is different: the authors in  \cite{LalegKirati15,LalegKirati16} start from a  continuous mathematical representation of the signal, and the discretization only occurs at the end of the process, giving a different basis set with different parameters to tune.  

In our proposed framework, both the energy of the wave functions and Planck's constant can be controlled, giving two parameters to tune ensuring flexibility in applications such as denoising. The method can be applied to any signal or image. In the presence of noise, quantum localization effects are present that may affect the denoising procedure, and we show how to avoid this problem by a suitable smoothing of the potential. 

The interest of the proposed adaptive transform is illustrated in this paper through a denoising application in the presence of signal or image-dependent noise. We show through specific examples that our method is more efficient than standard denoising through total variation regularized reconstruction methods.  We emphasize nevertheless that beyond these applications to specific signals, this paper also shows the interest of quantum mechanics techniques in signal and image processing in general and more particularly in constructing adaptive dictionaries. These adaptive basis could certainly find an interest in other applications than denoising and for other types of signal and images than those considered in this work.

The remainder of the paper is organized as follows. Section \ref{sec:quantum} presents a brief summary of the quantum mechanics principles used in this work. Section \ref{sec:adaptive} gives the details of the adaptive transform design and its application to denoising. Simulation results are provided in Section \ref{sec:results} and concluding remarks are finally reported in Section \ref{sec:conclusion}.

\section{Basics of quantum mechanics}
\label{sec:quantum}

In quantum mechanics, a (non-relativistic) particle in a potential is described by a wave function $\psi$, whose 
absolute value $|\psi |^2$ corresponds to the probability of presence of the particle. The normalization of probability implies that $\int |\psi |^2=1$, and the wave function belongs to the Hilbert space of functions with bounded integrals ($L^2$ norm). The wave function obeys the Schroedinger equation, whose stationary solutions for a given energy $E$ correspond to the equation:

\begin{equation}
 - \frac{\hbar ^2}{2m} \nabla ^2 \psi = - V(r)  \psi + E \psi
\label{eq:schrodinger}
\end{equation}

where $\hbar$ is the Planck constant and $m$ is the mass of the particle in the potential $V$.  It can also be written  $H \psi =  E \psi$ where $H$ is the Hamiltonian operator. If the space is discretized with a grid of $N$ points, \eqref{eq:schrodinger} has in general $N$ solutions which form a basis of the Hilbert space.

\section{Quantum signal and image processing}
\label{sec:adaptive}

\subsection{Adaptive transform design}

In order to apply these tools to signal processing, we replace in \eqref{eq:schrodinger} the potential $V$ by the image pixels' values (or the samples' values in the case of 1D signals). The proposed approach works in a discretized space, by replacing the Laplacian operator by its standard discrete version based on the numerical definitions of the gradient operator, given hereafter for an image $\bsx \in \mathbb{R}^{N\times N}$:

\begin{align}
(\partial_{\rm{h}} \bsx)(i,j) &= \bsx(i+1,j) - \bsx(i,j) \; &\textrm{if} \; i < N 
\nonumber
\\
(\partial_{\rm{v}} \bsx)(i,j) &= \bsx(i,j+1) - \bsx(i,j) \; &\textrm{if} \; j < N
\nonumber
\end{align}

where $\partial_{\rm{h}}$ and $\partial_{\rm{v}}$ are the horizontal and vertical gradients. Note that the boundary conditions considered for the gradients were a simple zero padding of the image.


The wave functions $\psi$ verifying \eqref{eq:schrodinger} can be numerically computed by calculating the eigenvectors of the Hamiltonian operator $H$. Given the discrete version of the Laplacian operator introduced above and replacing the potential $V$ by the pixels of image $\bsx$, the matrix $H \in \mathbb{R}^{N^{2} \times N^{2}}$ is defined as:

\begin{eqnarray}
\label{eq:H}
H(i,j)= \left \{
   \begin{array}{r c l}
      \bsx(i,j)+ 4 \frac{\hbar ^2}{2m} &  & for \; i=j,\\
       -\frac{\hbar ^2}{2m} & & for \; i = j \pm 1,\\
        -\frac{\hbar ^2}{2m} & & for \; i = j \pm N,\\
      0 & & otherwise.
   \end{array}
   \right.
\end{eqnarray}

As explained previously, zero padding is used to consider the boundary effects. As a consequence, specific coefficients of matrix $H$ do not follow the rule in \eqref{eq:H}. Specifically, $H(i,j) = \bsx(i,j)+ 2 \frac{\hbar ^2}{2m}$ for $i=j$ and $i \in \{1,...,N,N^2-N+1,...,N^2\}$, $H(i,j) = \bsx(i,j)+ 3 \frac{\hbar ^2}{2m}$ for any $i=j$ other than the previous set and $i \mod N^2 \in \{0,1\}$ and $H(i,i+1)=H(i+1,i)=0$ for any $i$ multiple of $N$.

The basis formed by the $N^2$ eigenvectors, denoted by $\bpsi_{i} \in \mathbb{R}^{N^2 \times 1}$, represents the adaptive transform proposed in this work. We consider in the following that these basis vectors are ordered from the highest to the lowest corresponding eigenvalues. The basis vectors $\bpsi_{i}$ are oscillating functions, similar to the Fourier or wavelet basis, but with a local frequency
given by the local value of $\sqrt{2m(E-V)}/\hbar$.  For a given value of energy $E$, wave vectors with energy below $E$ will use a larger range of frequencies to probe regions with low potential values than the ones corresponding to high potential values.
The precise relationship between $E-V$ and the local frequency depends on the parameter $\hbar^2/2m$, which in our case is a free parameter to be tuned. In quantum mechanics, it corresponds to the ratio of Planck's constant to the mass of the particle.  For very large values of $\hbar^2/2m$, the difference between high and low potential values becomes smaller and smaller, and the wave vectors basis becomes closer and closer to the usual Fourier basis.

\subsection{Application to denoising}

Most denoising methods consist in three steps: projecting a signal or an image onto a dictionary able to separate the useful information from the noise, followed by a thresholding (e.g., hard or soft) procedure in the transformed domain and finally by the recovery of the denoised signal or image by back projecting the modified coefficients into the time or space domain. In this work, we follow this  framework to show the interest of the adaptive basis formed by the basis vectors $\bpsi_{i}$ in denoising signals or images. As explained previously, the shape of the vectors $\bpsi_{i}$ depends on the value of $\frac{\hbar ^2}{2m}$, but is also influenced by the amount of noise degrading the signal or the image. More precisely, the basis vectors tend to spread out in time (or space) for slowly varying signals or images, and, on the contrary, contract in the presence of noise. This is due to a subtle quantum effect called Anderson localization (which justified a Nobel prize in physics in 1977): wave functions in a disordered (\textit{i.e.}, noisy in our case) system are in general exponentially localized in a specific location of the potential due to destructive interference \cite{Anderson1958}. To mitigate this effect, the Hamiltonian matrix is computed from a smoothed version of the noisy signal or image. This smooth version is computed in this work by a simple convolution with a Gaussian kernel whose standard deviation is denoted by $\sigma$. The denoised signal or image is finally reconstructed as follows:

\begin{eqnarray}
\label{eq:recons}
\hat{\bsx}= \sum _{i = 1} ^{N^2} \alpha_{i}\bpsi_{i}\tau_{i},
\end{eqnarray}

with 

\begin{eqnarray}
\label{eq:thr}
\tau_{i}= \left \{
   \begin{array}{r c l}
      1 &  & for \; i \leq s ,\\
      1 - \frac{i-s}{\rho} & & for \; i > s \;  and \; for \; 1 - \frac{i - s}{\rho} > 0 ,\\
      0 & & otherwise.
   \end{array}
   \right .
\end{eqnarray}

$\alpha_{i}$ are the coefficients representing the image $\bsx$ in the proposed adaptive basis. $s$ and $\rho$ are two hyperparameters that define the thresholding function used within our denoising algorithm. 

Algorithm \ref{Algo:Denoising} summarizes the steps of the denoising algorithm aiming at illustrating the practical interest of the proposed adaptive transform.

\begin{algorithm}[h!]
\label{Algo:Denoising}
\KwIn{$\bsx$, $\frac{\hbar ^2}{2m}$, $s$, $\rho$, $\sigma$}
\BlankLine
 {Compute a smooth version of $\bsx$ by Gaussian filtering}\\
 {Form the Hamiltonian matrix $H$ based on the smoothed version of $\bsx$} using \eqref{eq:H}\\
 {Calculate the eigenvectors $\bpsi_{i}$ of $H$}\\
 {Compute the coefficients $\alpha_{i}$ by projecting $\bsx$ onto the basis formed by $\bpsi_{i}$}\\
{Threshold the coefficients $\alpha_{i}$ and recover the denoised signal or image following \eqref{eq:thr} and \eqref{eq:recons}}\\
\KwOut{$\hat{\bsx}$}
\caption{Denoising algorithm using the proposed adaptive transform.}
\DecMargin{1em}
\end{algorithm}

\section{Simulation results}
\label{sec:results}

\begin{figure*}[h!]
  \centering
  \subfigure[]{\includegraphics[width=0.3\textwidth]{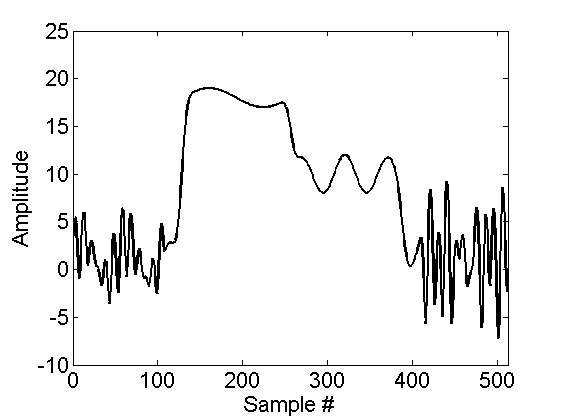}}
  \subfigure[]{\includegraphics[width=0.3\textwidth]{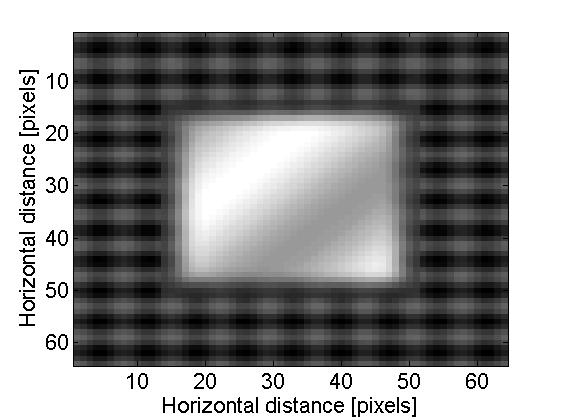}}
  \subfigure[]{\includegraphics[width=0.3\textwidth]{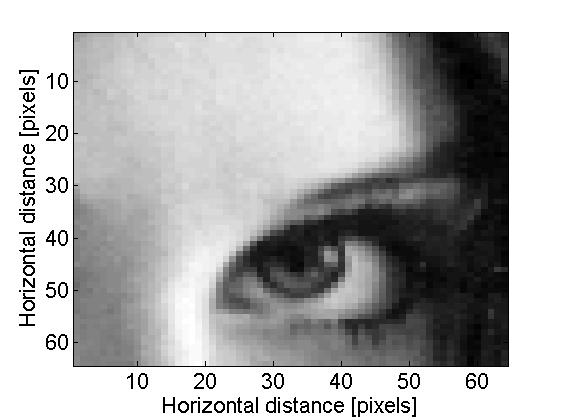}}
 \\
   \subfigure[]{\includegraphics[width=0.3\textwidth]{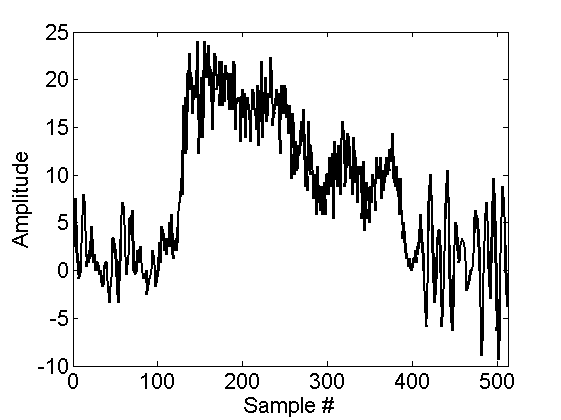}}
   \subfigure[]{\includegraphics[width=0.3\textwidth]{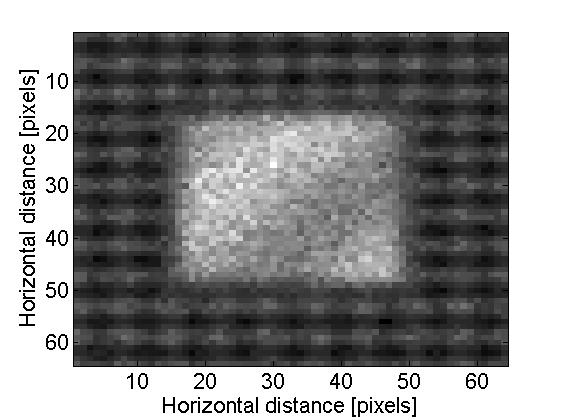}}
   \subfigure[]{\includegraphics[width=0.3\textwidth]{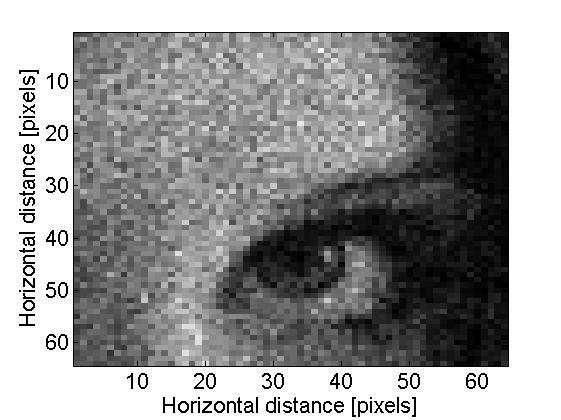}}
   \caption{(a-c) Clean signal and images, (d-f) Corresponding signal and images degraded by Poisson noise.}
\label{fig:SimulatedData}
\end{figure*}

In this section, we illustrate the proposed adaptive transform and its interest in denoising through three simulations: one synthetic signal, one synthetic image and a cropped version of Lena. In order to better emphasize the interest of the proposed framework, the synthetic signal and image are generated so that they contain low frequencies in the regions containing high samples or gray values, and reciprocally high frequencies for low potential regions. Two classes of simulations have been performed. Within the first class, the signal and the images were corrupted by Poisson noise. The samples' and pixels' values have been adjusted so that the SNR was as close as possible to $15$ dB. Note that in all the simulations, the SNR was computed by considering the noise as the difference between the noisy and clean signals or images. Within the second set of experiments, additive Gaussian noise was considered, with the variance pixel-wise (or sample-wise) related to the gray level or signal amplitude. The power of the noise was adjusted so that the signal to noise ratio (SNR) was $15$ dB. The noiseless and respectively noisy versions of these signal and images are shown in Fig. \ref{fig:SimulatedData}. Note that only the noisy and denoised signal and images corresponding to Poisson noise are shown herein for illustration purpose. The visual impression of the signal and images from the second experiment were very similar.

\begin{figure}[h!]
  \centering
\includegraphics[width=0.49\textwidth]{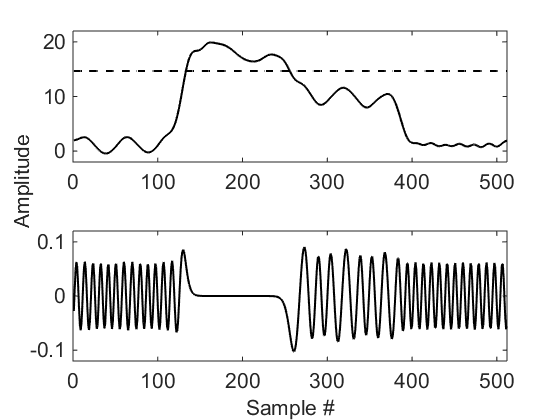}
   \caption{Up: blurred version of the noisy signal in Fig. \ref{fig:SimulatedData}(d), down: eigenvector corresponding to the eigenvalue illustrated by the dashed line.}
\label{fig:VP}
\end{figure}

An example of eigenvector extracted from the proposed adaptive basis is shown in Fig. \ref{fig:VP}. It has been computed, as suggested by Algorithm \ref{Algo:Denoising}, from a blurred version of the noisy signal in in Fig. \ref{fig:SimulatedData}(d). Its corresponding eigenvalue is illustrated by the dashed line superimposed to the blurred signal in Fig. \ref{fig:VP}. As expected, we remark that the lower the signal values, the higher the local frequency of the eignevector. Moreover, we observe that the signal regions larger than the eigenvalue are not captured by the eigenvector.

The main objective of our work was to introduce an original way of revisiting image/signal processing techniques with quantum physics tools, rather than finding a denoising method outperforming the state-of-the art methods in the field. However, the interest of the proposed adaptive basis in denoising was evaluated compared to an existing method formulating the denoising as an inverse problem. The latter, denoted as "TV", takes into account explicitly the Poisson noise and uses the total variation semi-norm to regularize the solution \cite{Figueiredo10}. For all the methods, the hyperparameters have been manually tuned to provide the best peak signal to noise ratios (PSNR). In addition to the PSNR, the structure similarity (SSIM) \cite{Wang04} and the SNR were used to evaluate the quality of denoised images compared to the ground truth. 

\begin{figure*}[h!]
  \centering
 \subfigure[]{\includegraphics[width=0.3\textwidth]{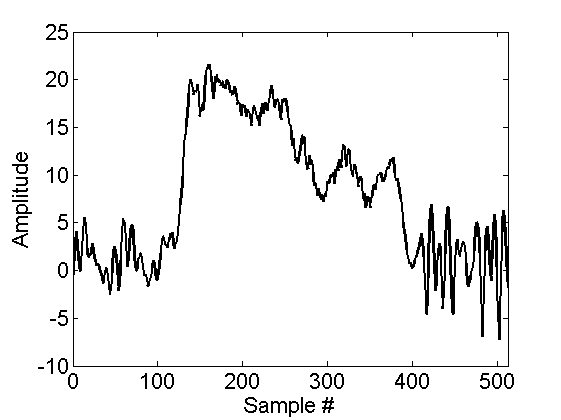}}
   \subfigure[]{\includegraphics[width=0.3\textwidth]{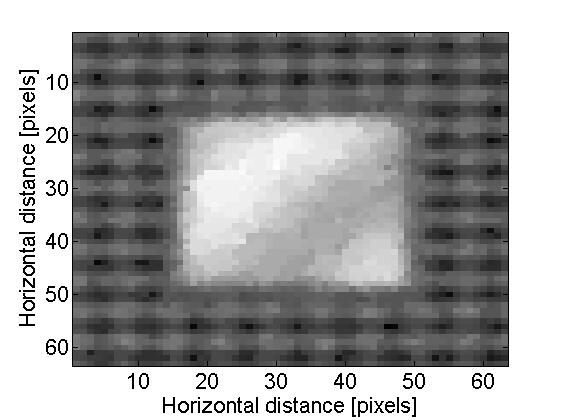}}
   \subfigure[]{\includegraphics[width=0.3\textwidth]{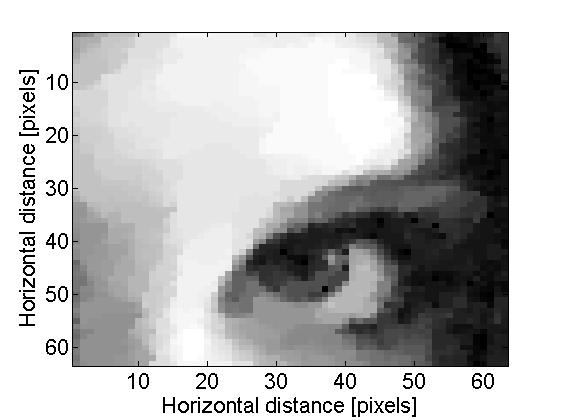}}
   \\
   \subfigure[]{\includegraphics[width=0.3\textwidth]{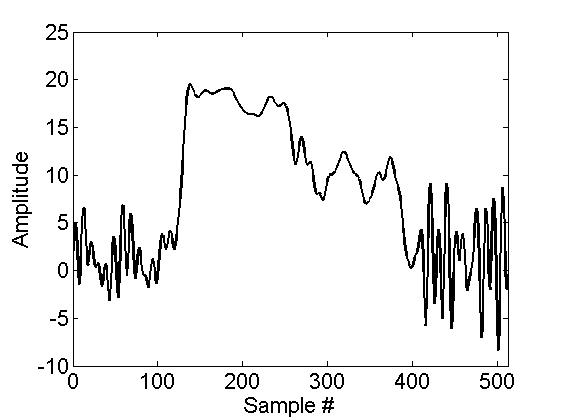}}
   \subfigure[]{\includegraphics[width=0.3\textwidth]{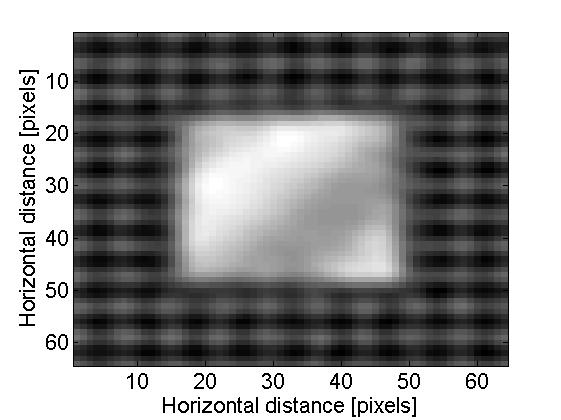}}
   \subfigure[]{\includegraphics[width=0.3\textwidth]{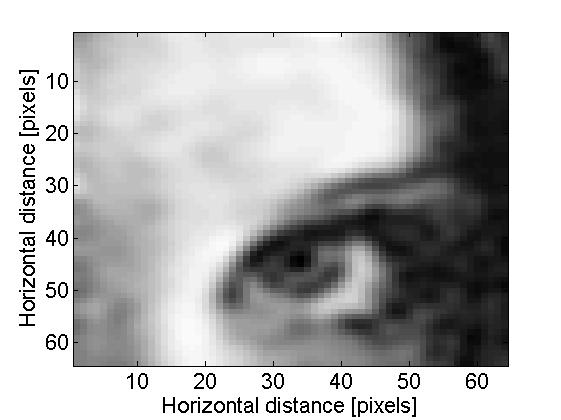}}
   
   \caption{(a-c) Denoising results for the Poisson noise experiments obtained (a-c) using total variation regularization and (d-f) using the proposed approach.}
\label{fig:Results}
\end{figure*}

The superiority of the denoising method based on thresholding the coefficients representing the noisy signal or image in the proposed adaptive basis can be qualitatively appreciated from the results in Fig. \ref{fig:Results} and quantitatively from the measures in Table \ref{tab_gp}. As expected, the total variation regularization provides reasonably good results for Lena image, but fails in recovering the synthetic signal and image because of their non piece-wise constant nature. In all the cases, although the data are not displayed in this paper, we checked that denoising using the proposed adaptive basis also outperformed classical filtering in the Fourier domain.

\begin{table}[h!]
\begin{center}
\caption{Quantitative denoising results.}
\label{tab_gp}
\begin{tabular}{|c|c|c|c|c|c|}
\hline
Data & Noise & Method  & PSNR (dB) & SNR (dB) & SSIM \\
 \hline
 \multirow{4}{*}{Signal}&
 \multirow{2}{*}{Poisson}
&TV    & 24.11 & 17.91 & NA\\
& &Proposed & 30.06 & 24.73 & NA \\
& \multirow{2}{*}{Gaussian}
&TV    & 23.52 & 17.44 & NA\\
& &Proposed & 27.80 & 22.69 & NA \\
\hline
 \multirow{4}{*}{Image}&
 \multirow{2}{*}{Poisson}
&TV    & 27.85 & 19.27 & 0.81\\
& &Proposed & 33.28 & 24.80 & 0.94 \\
& \multirow{2}{*}{Gaussian}
&TV    & 27.85 & 19.27 & 0.82\\
& &Proposed & 32.45 & 24.04 & 0.92 \\
\hline
 \multirow{4}{*}{Lena}&
 \multirow{2}{*}{Poisson}
&TV    & 27.98 & 23.98 & 0.80\\
& &Proposed & 28.69 & 24.70 & 0.82 \\
& \multirow{2}{*}{Gaussian}
&TV    & 27.07 & 22.53 & 0.69\\
& &Proposed & 32.45 & 24.04 & 0.92 \\

\hline
\end{tabular}
\end{center}
\end{table}

\vspace{-0.6cm}
\section{Conclusions}
\label{sec:conclusion}

This paper presented an original method of generating an adaptive transform for signal and image applications, inspired from concepts of quantum mechanics. More precisely, the proposed transform consists in the eigenvectors of the Hamiltonian operator. The interest of such an adaptive basis was shown through a denoising application in presence of Poisson noise. 
As a perspective of this work, an interesting research track is to extend the study to other signal and image reconstruction applications, such as deconvolution or super-resolution. Moreover, the computational time could be mitigated by an efficient method of computing the eigenvectors from the Hamiltonian matrix.

\clearpage
\bibliographystyle{IEEEbib}
\bibliography{biblio}

\end{document}